\documentclass[twocolumn,english,aps,prb,twocolumn,superscriptaddress,natbib,bibnotes,amsmath,amssymb,floatfix,groupedaddress,footinbib]{revtex4-2}

\usepackage[colorlinks=true,citecolor=blue,linkcolor=magenta]{hyperref}

\usepackage[markup=nocolor, authormarkupposition=left]{changes} 

\usepackage{soul}
\usepackage[utf8]{inputenc}
\usepackage[english]{babel}
\usepackage{amsmath,amsfonts,amssymb}
\usepackage[T1]{fontenc}
\usepackage{url}

\usepackage{amsmath}
\usepackage{siunitx}
\usepackage[version=4]{mhchem}
\usepackage{amsfonts}
\usepackage{amssymb}

\usepackage{epstopdf}
\usepackage{graphicx}
\graphicspath{{./Figures/}}

\usepackage{changes}

\begin{document}
\title{Programmable generation of optical skyrmions on a silicon photonic chip}

\author{Mingyuan Zhang$^{1,*}$, Xiaofu Pan$^{2,*}$, Wu Zhou$^{1,*}$, Wenzhang Tian$^{1}$, Zengqi Chen$^{1}$, Yiou Cui$^{1}$, Yijie Shen$^{3,4}$, Yeyu Tong$^{1,\dagger}$, \& Jianqi Hu$^{2,\ddagger}$}

\affiliation{
$^1$Microelectronic Thrust\char`,{} The Hong Kong University of Science and Technology (Guangzhou)\char`,{} 511453\char`,{} Guangzhou\char`,{} Guangdong\char`,{} China \\
$^2$Department of Electrical and Computer Engineering\char`,{} The University of Hong Kong\char`,{} Hong Kong\char`,{} China\\
$^3$Centre for Disruptive Photonic Technologies\char`,{} School of Physical and Mathematical Sciences\char`,{} Nanyang Technological University\char`,{} Singapore 637371\char`,{} Singapore\\
$^4$School of Electrical and Electronic Engineering\char`,{} Nanyang Technological University\char`,{} Singapore 639798\char`,{} Singapore\\
}

\maketitle

\noindent\textbf{\noindent
Optical skyrmions, characterized by topologically stable and spatially varying polarization textures, show immense potential for robust optical communications and metrology. However, conventional methods for generating optical Stokes skyrmions rely on bulky free-space optics, strictly constraining both system miniaturization and dynamic reconfigurability. Here, we demonstrate the efficient and programmable generation of optical skyrmions and bimerons using a compact silicon photonic chip. By integrating a programmable Mach--Zehnder interferometer mesh with a multi-dimensional grating emitter, we dynamically control the amplitudes, phases, and polarizations of emitted fundamental and orbital angular momentum modes. This architecture allows on-demand electrical switching among a complete library of optical quasi-particle states, including N\'eel, Bloch, intermediate, and anti-type skyrmions and bimerons. Experimental full-Stokes polarimetry confirms high-fidelity polarization textures with near-unity skyrmion numbers. Our foundry-compatible platform translates complex topological light generation into simple voltage controls, paving the way for next-generation communication and sensing systems based on optical skyrmions.}

\section*{Introduction}

\noindent{Skyrmions} are topologically stable, knot-like field configurations characterized by an integer winding number, also known as the Skyrmion number. Originally introduced by Tony Skyrme in nuclear physics \cite{skyrme1961non}, these non-trivial textures have since been observed and realized across a broad spectrum of physical systems. Notable examples include Bose--Einstein condensates \cite{al2001skyrmions}, magnetic materials  \cite{nagaosa2013topological}, liquid crystals \cite{foster2019two,duzgun2021skyrmion}, twistronics \cite{khalaf2021charged}, water waves \cite{wang2025topological}, acoustics \cite{ge2021observation}, and optics \cite{shen2024optical}. In the optical domain, skyrmions can be engineered through various vector field representations \cite{cheng2025navigating}, yielding diverse manifestations such as optical field \cite{tsesses2018optical}, spin \cite{du2019deep}, and Stokes \cite{gao2019paraxial} skyrmions. Beyond serving as a platform to explore fundamental physics of quasi-particles \cite{dai2020plasmonic,davis2020ultrafast}, optical skyrmions hold promise for next-generation applications in robust optical transport \cite{wang2024topological,nape2022revealing}, communications \cite{peters2025seeing}, optical manipulation \cite{wang2012optical}, optical storage \cite{Zhang2026Skyrmions}, information processing \cite{wang2025perturbation}, as well as advanced sensing and metrology \cite{yang2023spin,ma2025using}.

In the context of optical Stokes skyrmions, the underlying vector field is defined by the spatially varying, normalized Stokes vector of light \cite{gao2019paraxial,sugic2021particle,shen2022generation}. By intertwining spatial structures and polarization degrees of freedom, these optical states exhibit a complete mapping onto the polarization Poincar\'e sphere \cite{beckley2010full}. Conventionally, generating such optical skyrmions relies on bulk polarization optics and spatial light modulators \cite{sugic2021particle,shen2022generation}, graded-index optical elements \cite{shen2024topologically}, and metafibres \cite{he2024optical}. While these approaches offer flexible wavefront control, they typically require complex free-space optical setups and careful alignment, and provide limited dynamic reconfigurability. More recently, nanophotonic and photonic integrated platforms have enabled the generation of optical skyrmions within compact footprints using microcavities \cite{lin2021microcavity,lin2024chip,lai2026chip,wei2026nanophotonic}, metasurfaces \cite{li2024realization,liu2026chip}, and structured materials \cite{ma2025nanophotonic,liu2026broadband}. However, the skyrmionic field textures in most of these devices are rigidly defined by the fabricated structures or resonant modes. Dynamic reconfiguration among multiple skyrmion textures, and between skyrmion and bimeron families \cite{shen2021topological}, remains comparatively underexplored on a single photonic chip.

Programmable photonics offers a natural route to overcome this limitation \cite{bogaerts2020programmable,carolan2015universal}. By employing integrated interferometric meshes with tunable phase shifters, programmable photonic integrated circuits have driven significant advances in fields such as optical signal processing \cite{perez2017multipurpose}, microwave photonics \cite{perez2024general}, quantum optics \cite{harris2017quantum}, and photonic computing \cite{shen2017deep}. Furthermore, when integrated with versatile chip-to-free-space interfaces, these platforms provide programmable control over the amplitude, phase, polarization, and spatial profiles, including orbital angular momentum (OAM), of the emitted light fields \cite{zhou2024ultra,zheng2026fully,valdez2026integrated}, enabling on-chip structured light generation \cite{butow2024generating} and multi-mode optical communications \cite{seyedinnavadeh2024determining,lu2024empowering}.

\begin{figure*}[t]
\centering
\includegraphics[width=\textwidth]{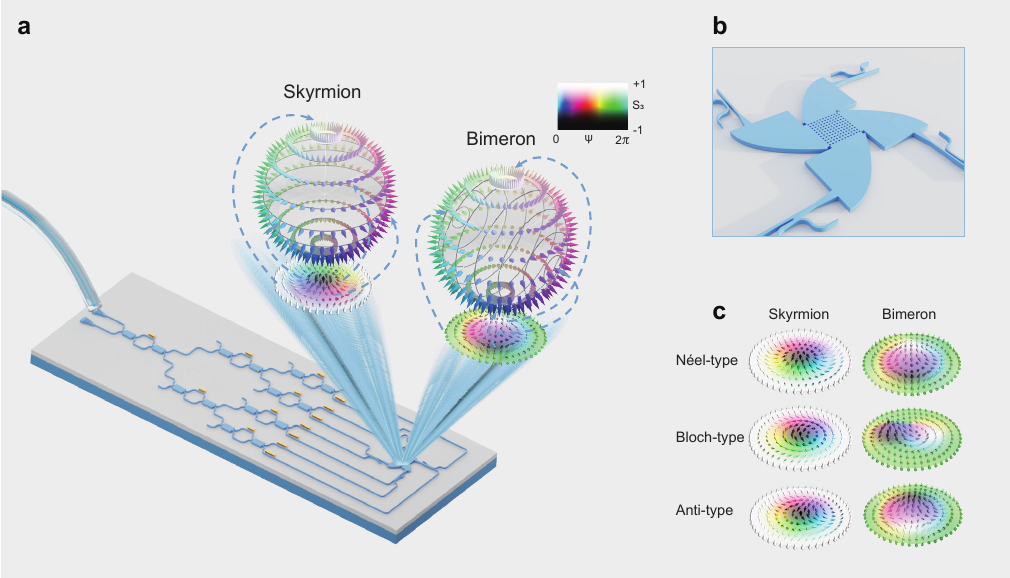}
\caption{\textbf{Schematic of the programmable generation of optical skyrmions and bimerons on chip.} \textbf{a}, Artistic illustration of generating optical quasi-particle states on a programmable silicon photonic chip. The chip consists of a one-by-eight, binary-tree programmable Mach--Zehnder interferometer (MZI) mesh and a multi-dimensional emitter based on a space- and polarization-diversity grating coupler. By coupling laser light via a grating coupler and controlling the relative amplitudes and phases of eight fundamental transverse-electric (TE) modes injected into the emitter, the diffracted beam structures in free-space can be programmed to generate versatile optical skyrmions and bimerons. The normalized Stokes fields are mapped onto the Poincar\'e sphere, where colour denotes the polarization azimuth \(\psi\) and lightness represents the normalized Stokes component \(S_3\). \textbf{b}, Enlarged schematic of the multi-dimensional emitter, comprising a two-dimensional grating region on silicon with shallowly etched holes, four parabolic spot-size converters (SSCs), adiabatic directional couplers (ADCs), and eight input waveguides. \textbf{c}, Calculated polarization textures of N\'eel-, Bloch- and anti-type optical skyrmion and bimeron states. Arrows show the in-plane Stokes-vector components, and the background colour represents the local polarization state.}
\label{fig:1}
\end{figure*}

In this work, we leverage the principle of programmable photonics to generate reconfigurable optical skyrmions on a silicon photonic chip. The chip integrates a binary-tree Mach--Zehnder interferometer (MZI) mesh with a multi-dimensional grating emitter \cite{zhou2024ultra} to program optical Stokes textures. This architecture enables highly efficient on-chip generation of optical skyrmions, spanning characteristic states from N\'eel-I, Bloch, intermediate, to N\'eel-II and anti-skyrmion types. Moreover, the same device can be flexibly reprogrammed to generate representative classes of optical bimerons. We experimentally characterize the generated optical quasi-particle states using Stokes polarimetry, where the measured polarization textures closely agree with theory and yield absolute topological numbers close to unity. With calibrated voltage settings, our approach unlocks straightforward, on-demand optical skyrmion generation on a compact, foundry-compatible silicon photonic platform.

\section*{Results}
\noindent\textbf{Principle of operation.} An optical Stokes skyrmion can be formed by coherently superimposing a fundamental Gaussian beam and an OAM-carrying Laguerre--Gaussian (LG) beam with orthogonal polarization states \cite{gao2019paraxial}. For the states considered here, the vector field can be written as
\begin{equation}
\label{eq:superposition}
\left|\Psi(\mathbf{r})\right\rangle
=
a_{0}\,\mathrm{LG}_{0,0}(\mathbf{r})\left|e_1\right\rangle
+
a_{\ell}e^{-i\Delta\varphi}\mathrm{LG}_{0,\ell}(\mathbf{r})\left|e_2\right\rangle .
\end{equation}
Here, \(\mathrm{LG}_{0,0}(\mathbf{r})\) is the fundamental Gaussian mode, \(\mathrm{LG}_{0,\ell}(\mathbf{r})\) is the first-order OAM mode with \(\ell=\pm1\), and \(a_0\) and \(a_\ell\) are their respective amplitudes. The kets \(\left|e_1\right\rangle\) and \(\left|e_2\right\rangle\) denote the polarization states, and \(\Delta\varphi\) is the relative phase between the two modal components. The superposition produces a spatially varying polarization texture described by the normalized Stokes vector,
\[
\mathbf{S}(\mathbf{r})=
\left[S_1(\mathbf{r}),S_2(\mathbf{r}),S_3(\mathbf{r})\right]
=
\frac{\left[s_1(\mathbf{r}),s_2(\mathbf{r}),s_3(\mathbf{r})\right]}{s_0(\mathbf{r})}.
\]
For the evaluation of the skyrmion number, the measured Stokes vector is further normalized to unit length as \(\mathbf{n}(\mathbf{r})=\mathbf{S}(\mathbf{r})/|\mathbf{S}(\mathbf{r})|\).

\begin{figure*}[t]
\centering
\includegraphics[width=\textwidth]{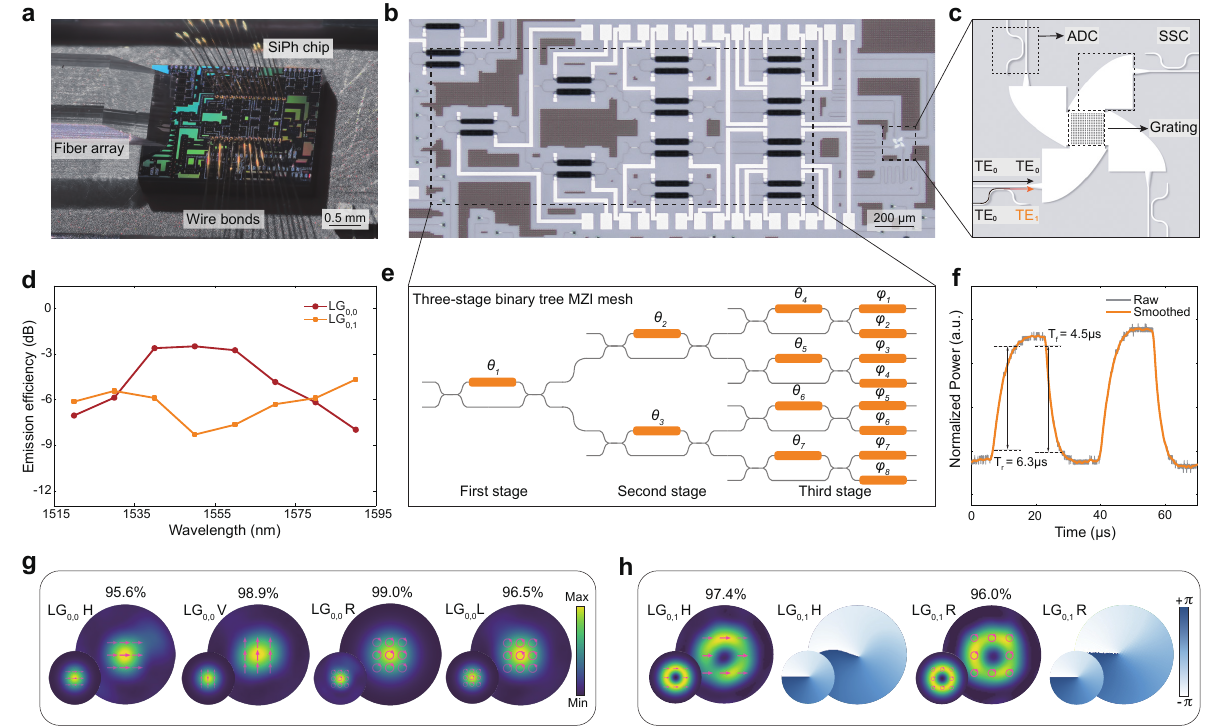}
\caption{\textbf{Design, characterization, and calibration of the programmable photonic chip.} \textbf{a}, Photograph of the packaged silicon photonic chip with fibre-array coupling and electrical wire bonding. \textbf{b}, Optical micrograph of the photonic integrated circuit, showing the binary-tree MZI mesh and the multi-dimensional emitter. \textbf{c}, Schematic of the multi-dimensional emitter region consisting of a two-dimensional grating, SSCs and ADCs, with the \(\mathrm{TE}_0\) and \(\mathrm{TE}_1\) input modes indicated. \textbf{d}, Measured wavelength-dependent emission efficiencies of the \(\mathrm{LG}_{0,0}\) and \(\mathrm{LG}_{0,1}\) modes. \textbf{e}, Schematic of the three-stage binary-tree MZI mesh. The phase shifters \(\theta_1\)--\(\theta_7\), distributed across the three splitting stages, control the relative output amplitudes. \(\varphi_1\)--\(\varphi_8\) independently tune the phases of the eight outputs. \textbf{f}, Measured temporal response of a representative heater-actuated MZI. The raw and smoothed traces show rise and fall times (\(10\%\) to \(90\%\)) of \(6.3~\mu\mathrm{s}\) and \(4.5~\mu\mathrm{s}\), respectively. \textbf{g}, Simulated (left insets) and measured far-field intensity distributions of the generated \(\mathrm{LG}_{0,0}\) mode in the H, V, R, and L polarization states. The fidelity shown is the Pearson correlation coefficient between the measured and theoretical intensity distributions. \textbf{h}, Simulated (left insets) and measured intensity and phase distributions of representative \(\mathrm{LG}_{0,1}\) modes in the H and R polarization states. The fidelity shown is defined in the same manner as in \textbf{g}.}
\label{fig:2}
\end{figure*}

Figure~\ref{fig:1}a illustrates the principle for programmable generation of these optical quasi-particle states on a silicon photonic chip. The chip integrates a programmable MZI mesh tuned by thermal heaters with a multi-dimensional emitter based on grating couplers \cite{zhou2024ultra,tong2019efficient} (Fig.~\ref{fig:1}b). Within the supported mode subspace, the MZI mesh accurately configures the relative amplitudes and phases of eight waveguide modes directed into the grating region, which are subsequently diffracted and transformed into free-space propagating modes with OAM orders of \(\ell=0,\pm1\) and arbitrary polarization states. Here, circular polarization mode bases (right-handed/left-handed, denoted as R/L) are utilized to construct the skyrmion family, whereas linear polarization mode bases (horizontal/vertical, denoted as H/V) generate the bimeron family. By mapping the spatial coordinates of the emitted beam to points on the Poincar\'e sphere, various kinds of optical quasi-particles can be synthesized. The hue-lightness (HL) map visualizes the local polarization state, with hue encoding the azimuthal coordinate on the Poincar\'e sphere and lightness encoding the Stokes parameter \(S_3\). For optical skyrmions, the Stokes vector covers the Poincar\'e sphere from one pole to the opposite pole as the radial coordinate increases from the beam centre to the edge. A global rotation of the Stokes field converts this polar configuration into a bimeron texture, where the Stokes vector instead evolves between two antipodal points on the equator of the Poincar\'e sphere \cite{shen2021topological}. Adjusting the relative phases between different waveguide modes controls the helicity of the texture, while reversing the relative OAM winding switches between the conventional and anti-type states (Fig.~\ref{fig:1}c).

\begin{figure*}[t]
\centering
\includegraphics[width=0.9\textwidth]{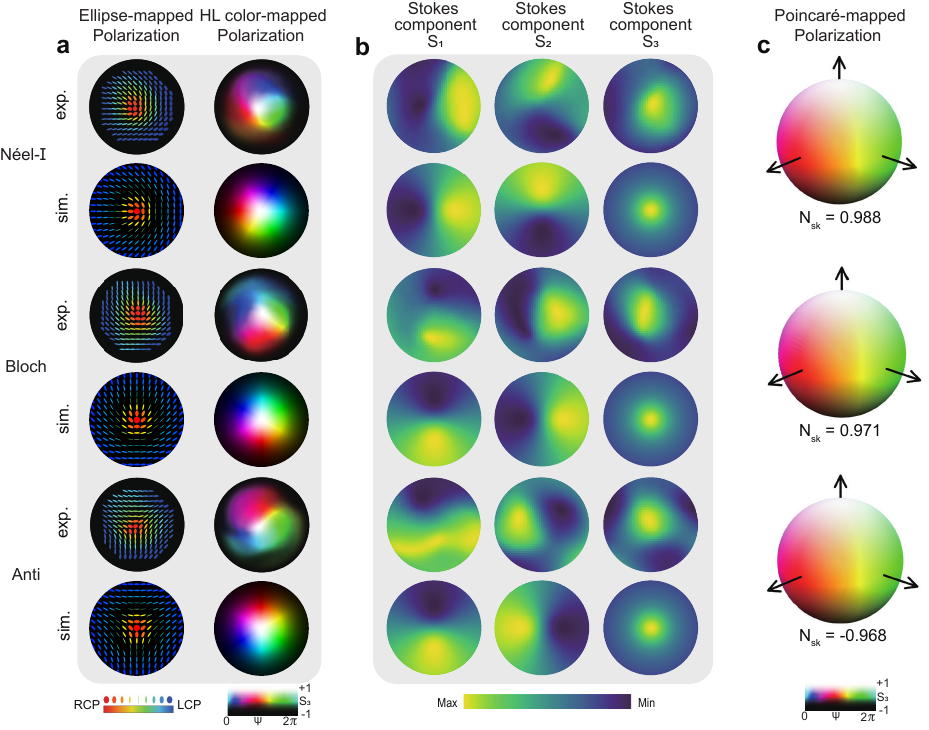}
\caption{\textbf{Generation and characterization of reconfigurable optical skyrmions.} \textbf{a}, Measured (exp.) and simulated (sim.) ellipse-mapped and HL colour-mapped polarization distributions for N\'eel-I, Bloch and anti-type skyrmions. \textbf{b}, Corresponding spatial distributions of the Stokes components \(S_1\), \(S_2\) and \(S_3\). \textbf{c}, Polarization textures of generated skyrmions mapped onto the Poincar\'e sphere. The reconstructed topological numbers are \(N_{\mathrm{sk}}=0.988\), \(0.971\) and \(-0.968\) for the N\'eel-I, Bloch, and anti-type optical skyrmions, respectively.}
\label{fig:3}
\end{figure*}

\vspace{1mm}
\noindent\textbf{Programmable photonic chip.} The silicon photonic chip we use to generate optical skyrmions is optically packaged and electrically wire bonded to facilitate the experimental measurements (Fig.~\ref{fig:2}a). As shown in Fig.~\ref{fig:2}b, a programmable MZI mesh and a multi-dimensional emitter are integrated within a compact footprint of \(2.8~\mathrm{mm}\times1.1~\mathrm{mm}\). The three-stage programmable section employs seven heaters within MZIs to control the relative amplitudes of the eight output waveguides, followed by an additional set of heaters that independently tune their phases (Fig.~\ref{fig:2}e). These outputs are subsequently routed to the emitter section. The multi-dimensional emitter comprises three key components (Fig.~\ref{fig:2}c), including adiabatic directional couplers (ADCs) for \(\mathrm{TE}_0\) and \(\mathrm{TE}_1\) mode multiplexing \cite{ding2013chip_OE}, spot-size converters (SSCs) for mode-field expansion \cite{xu2023compact}, and a four-port two-dimensional grating that diffracts the waveguide modes into free-space propagating modes.

By tuning the relative amplitudes and phases of these waveguide modes, fundamental \(\mathrm{LG}_{0,0}\) and \(\mathrm{LG}_{0,\pm1}\) modes of light can be vertically emitted from the photonic chip (see Supplementary Note~1). Their emission efficiencies are wavelength-dependent, which are experimentally characterized (Fig.~\ref{fig:2}d) and align well with the numerical simulations (see Supplementary Note~1). Near the \(1580~\mathrm{nm}\) operating wavelength, the measured emission efficiencies of both the \(\mathrm{LG}_{0,0}\) and \(\mathrm{LG}_{0,1}\) modes are approximately \(-6~\mathrm{dB}\). Without loss of generality, we synthesize optical quasi-particle states at this wavelength, while the scheme works virtually at any wavelength across the C-band by compensating for the efficiency imbalance.

\begin{figure*}[t]
\centering
\includegraphics[width=0.9\textwidth]{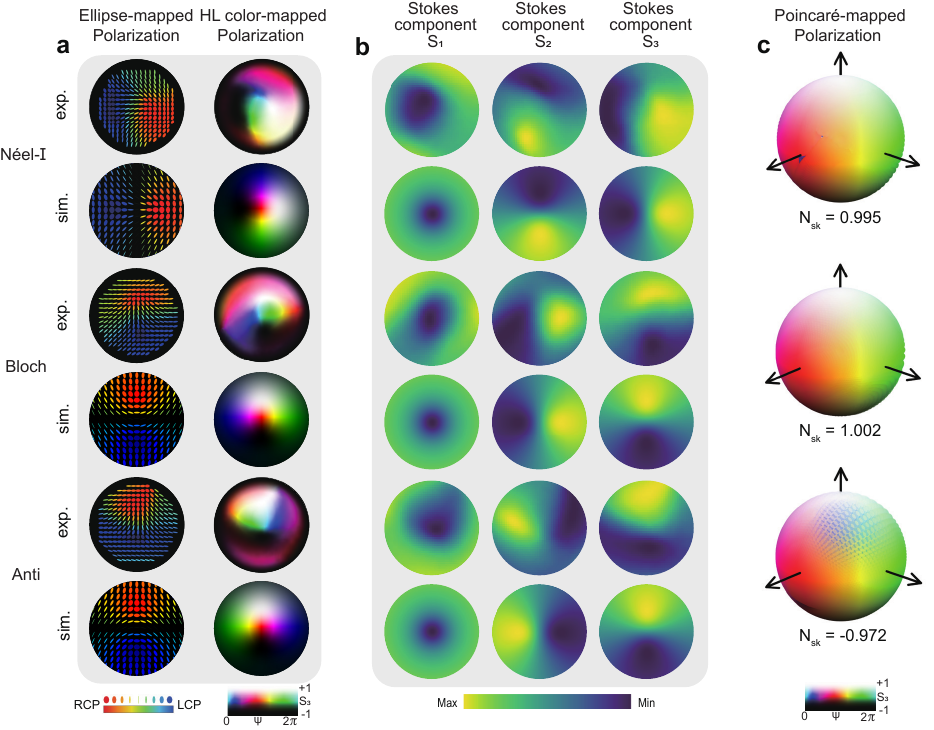}
\caption{\textbf{Generation and characterization of reconfigurable optical bimerons.} \textbf{a}, Measured (exp.) and simulated (sim.) ellipse-mapped and HL colour-mapped polarization distributions for N\'eel-I, Bloch and anti-type bimerons. \textbf{b}, Corresponding spatial distributions of the Stokes components \(S_1\), \(S_2\) and \(S_3\). \textbf{c}, Polarization textures of generated bimerons mapped onto the Poincar\'e sphere. The reconstructed topological numbers are \(N_{\mathrm{sk}}=0.995\), \(1.002\) and \(-0.972\) for the N\'eel-I, Bloch, and anti-type optical bimerons, respectively.}
\label{fig:4}
\end{figure*}

Before generating optical quasi-particle states on chip, we first calibrate the MZI mesh to establish a complete mapping from heater settings to the complex coefficients of the eight waveguide channels impinging on the multi-dimensional emitter. These heaters have typical switching times on the microsecond scale (Fig.~\ref{fig:2}f). Details of the heater characterization, as well as the amplitude and phase calibration of the programmable section, are provided in Supplementary Note~2. Following calibration, we synthesize the constituting components of the optical skyrmions and bimeron beams, i.e., \(\mathrm{LG}_{0,0}\) and \(\mathrm{LG}_{0,1}\) beams with R/L and H/V polarizations. Figures~\ref{fig:2}g and \ref{fig:2}h illustrate their experimental and simulated intensity distributions, alongside the phase profiles of representative \(\mathrm{LG}_{0,1}\) beams. To quantify the fidelity of LG beam generation, we calculate the Pearson correlation coefficient between the measured and theoretical spatial intensity distributions scaled to the same \(1/e^2\) beam radii (see Supplementary Note~3). Measured intensity fidelities range from \(95.6\%\) to \(99.0\%\), confirming that the emitter reproduces the desired free-space modal profiles with high accuracy. The OAM phase profiles of \(\mathrm{LG}_{0,1}\) beams are retrieved from the off-axis holography as described in Supplementary Note~4.

\vspace{1mm}
\noindent\textbf{Reconfigurable optical skyrmions and bimerons.} Optical skyrmions and bimerons can be formed by the coherent superposition of \(\mathrm{LG}_{0,0}\) and \(\mathrm{LG}_{0,\pm1}\) beams shown in Figs.~\ref{fig:2}g and \ref{fig:2}h (see Supplementary Note~2). With a pre-calibrated MZI mesh, these states can be synthesized by programming the heaters to obtain the required modal amplitudes and relative phases, thereby defining the spatially varying polarization texture of the emitted fields. For the skyrmion family, the two modal components are expressed in the circular-polarization (R/L) basis. Their relative phase \(\Delta\varphi\) controls the helicity of the skyrmion texture, with \(\Delta\varphi=0\), \(\pi/2\), and \(\pi\) generating the N\'eel-I, Bloch, and N\'eel-II states, respectively. The intermediate phases continuously rotate the in-plane Stokes-vector orientation, while reversing the OAM order generates the corresponding anti-type states.

Figure~\ref{fig:3} showcases the experimentally measured N\'eel-I, Bloch and anti-type optical skyrmions generated on chip, which are in good agreement with their ideal polarization textures. Here, the ellipse maps show the local polarization ellipses, while the HL maps illustrate a continuous representation of the normalized Stokes field. In the experiment, these polarization textures are characterized by spatially resolved full-Stokes polarimetry using a linear polarizer and a quarter-wave plate (see Supplementary Note~5). Figure~\ref{fig:3}b depicts the measured \(S_1\), \(S_2\) and \(S_3\) components that reproduce the main features of the simulations, while Fig.~\ref{fig:3}c illustrates the normalized Stokes vectors mapped onto the Poincar\'e sphere. Further, we calculate the skyrmion numbers of the synthesized beams based on the unit Stokes-vector field:
\begin{equation}
\label{eq:nsk}
N_{\mathrm{sk}}
=
\frac{1}{4\pi}
\iint_{\Omega}
\mathbf{n}\cdot
\left(
\frac{\partial\mathbf{n}}{\partial x}
\times
\frac{\partial\mathbf{n}}{\partial y}
\right)
\,dx\,dy .
\end{equation}
Here, \(\mathbf{n}=\mathbf{S}/|\mathbf{S}|\) and \(\Omega\) is the valid beam region used in the numerical calculation. The reconstructed skyrmion numbers are \(0.988\), \(0.971\) and \(-0.968\) for the N\'eel-I, Bloch and anti-type states, respectively, consistent with their designed skyrmion numbers. Characterization of other representative states, including intermediate and N\'eel-II optical skyrmions, is provided in Supplementary Note~6.

For optical bimerons, switching the polarization basis from R/L to H/V in Eq.~\eqref{eq:superposition} applies a global rotation to the Stokes field on the Poincar\'e sphere, thereby transforming the optical skyrmion into bimeron generation. Figure~\ref{fig:4} illustrates the N\'eel-I, Bloch, and anti-type optical bimerons emitted from the programmable silicon photonic chip. The experimentally reconstructed ellipse maps, HL maps and normalized Stokes components reproduce the corresponding simulated polarization textures. Their mappings onto the Poincar\'e sphere retain the expected topological coverage, with reconstructed skyrmion numbers of \(0.995\), \(1.002\), and \(-0.972\), respectively. The characterization of intermediate and N\'eel-II bimeron states is provided in Supplementary Note~6.

\section*{Discussion}

\begin{figure}[t]
\centering
\includegraphics[width=0.5\textwidth]{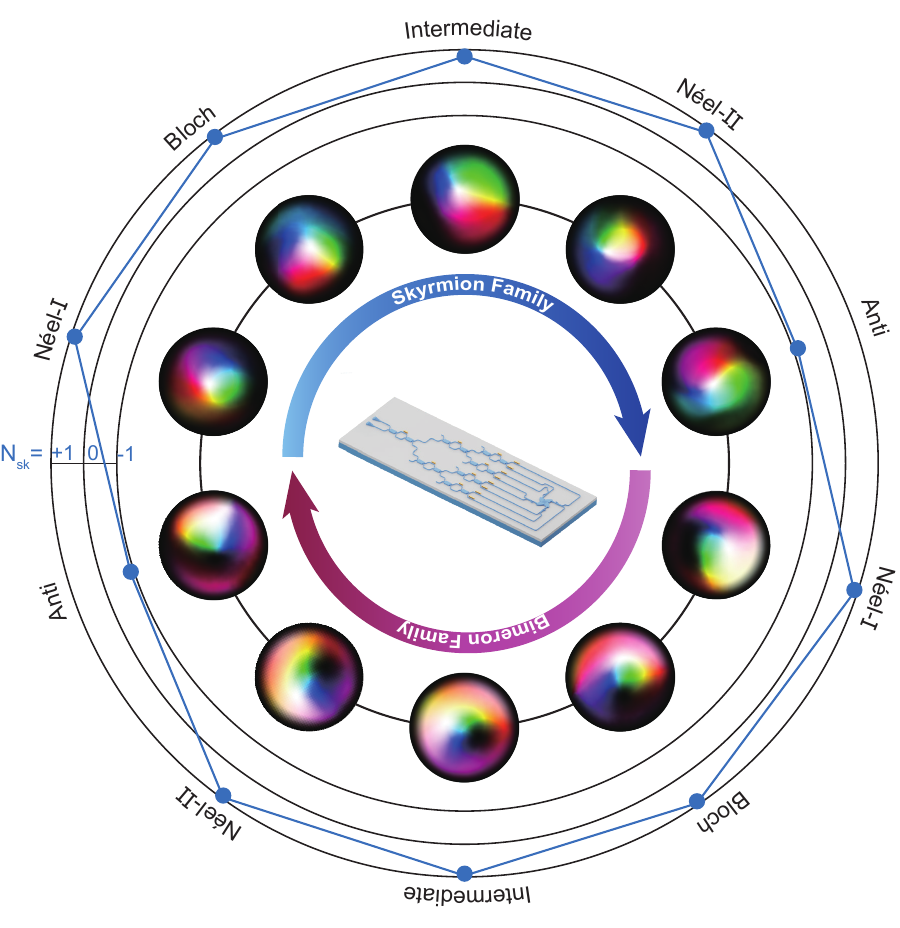}
\caption{\textbf{Programmable skyrmion and bimeron states on chip.} HL colour-mapped polarization distributions of representative skyrmion and bimeron states generated by the silicon photonic chip. The inner arrows indicate reconfiguration within the skyrmion and bimeron families, while the outer circles summarize the reconstructed skyrmion number \(N_{\mathrm{sk}}\) of each state.}
\label{fig:state_space}
\end{figure}

\noindent{In} this work, we demonstrate efficient and programmable generation of optical quasi-particle states on a compact optical chip. As summarized in Fig.~\ref{fig:state_space}, this silicon photonic chip can synthesize all representative types of optical skyrmions and bimerons ranging from the N\'eel-I, Bloch, intermediate, N\'eel-II, and anti-type states. These states are obtained by varying the polarization basis and the relative phase between \(\mathrm{LG}_{0,0}\) and \(\mathrm{LG}_{0,1}\) modes, as well as by reversing the OAM winding. Their measured skyrmion numbers all remain close to the target values, which are described in Figs.~\ref{fig:3}, \ref{fig:4} and Supplementary Note~6.

Thermo-optic phase shifters are employed in this study to reconfigure the polarization textures, offering compact footprint and microsecond switching speeds. Here, each heater operates within a \(0\)--\(4~\mathrm{mW}\) power range, and a \(2\pi\) phase shift requires approximately \(2.6~\mathrm{mW}\) for the representative phase-control heaters (see Supplementary Note~2). For the 15-heater programmable mesh used in our chip, the conservative upper bound of power consumption is approximately \(60~\mathrm{mW}\). The actual power consumption is state dependent and generally much lower, as the programmed states do not require all heaters to operate at their maximum power. Once calibrated, target optical quasi-particle states can be loaded directly via predetermined heater settings.

While offering better energy efficiency and faster switching speed than typical free-space systems, our current implementation remains constrained by the thermo-optic effect. Migrating this programmable architecture to electro-optic platforms, such as thin-film lithium niobate or lithium tantalate \cite{wang2018integrated,wang2024lithium}, will further increase reconfiguration speed and power efficiency. This can unlock the robust, gigahertz-rate optical communications based on optical skyrmions \cite{peters2025seeing,chen2026programmable}. Limited by the supported modes of the designed emitter, our chip can only synthesize optical quasi-particle states with the absolute skyrmion number \(\lvert N_{\mathrm{sk}}\rvert\) of 1. Extending the emitter to support additional spatial modes  \cite{sharma2025universal}, for example through inverse design \cite{zheng2026fully}, could enable higher-order skyrmion numbers and more complex topological textures.

Beyond footprint reduction, miniaturizing skyrmion generators overcomes the instability and alignment issues of traditional free-space setups, while facilitating their integration with other photonic functionalities and scalable fabrication. Crucially, introducing the concept of programmable photonics to skyrmion generation addresses key challenges in flexibly reconfiguring the polarity, vorticity, and helicity of optical skyrmions on chip. Ultimately, this architecture translates the topology of emitted light into electronically controllable voltages, paving the way for adaptive topological-state encoding, reconfigurable light--matter interactions, and fully integrated systems for next-generation optical communications and sensing.

\vspace{0.1cm}
\begin{footnotesize}
\textbf{Acknowledgments}:
Y.T. acknowledges support from the National Natural Science Foundation of China (No. 62305277) and the Guangdong Provincial Talent Program (No. 2024TQ08X295). J.H. acknowledges support from National Natural Science Foundation of China (No. 62622517) and the Early Career Scheme (Grant No. 27203626) from the Research Grant Council of Hong Kong. Y.S. acknowledges Singapore Ministry of Education (MOE) (RG157/23 \& RT11/23), Singapore Agency for Science, Technology and Research (A*STAR) (M24N7c0080 \& H25-MRO3489), and Nanyang Assistant Professorship Start Up grant. The authors acknowledge Jin Liu for helpful discussions.

\vspace{0.1cm}

\noindent \textbf{Data Availability Statement}: 
The data and code that support the plots within this paper and other findings of this study are available from the corresponding author upon reasonable request.

\noindent \textbf{Competing interests}: The authors declare no competing interests.
\end{footnotesize}

\renewcommand{\bibpreamble}{
$^*$These authors contributed equally to this work.\\
$^\dagger${Corresponding author: \textcolor{magenta}{yeyutong@hkust-gz.edu.cn}}\\
$^\ddagger${Corresponding author: \textcolor{magenta}{jianqi@hku.hk}}\\
}

\bibliographystyle{naturemag}
\bibliography{ref}

\end{document}